\documentclass[prl,groupedaddress]{revtex4}
\usepackage{epsf}
\usepackage[dvips]{graphicx}

\newcommand{\beq}{\begin{equation}}
\newcommand{\eeq}{\end{equation}}
\newcommand{\beqa}{\begin{eqnarray}}
\newcommand{\eeqa}{\end{eqnarray}}

\newcommand{\laem}{\begin{array}{c} < \vspace{-0.7em} \\ {\scriptstyle \sim}
\end{array}}
\newcommand{\gaem}{\begin{array}{c} > \vspace{-0.7em} \\ {\scriptstyle \sim}
\end{array}}

\def\slashchar#1{\setbox0=\hbox{$#1$}           
   \dimen0=\wd0                                 
   \setbox1=\hbox{/} \dimen1=\wd1               
   \ifdim\dimen0>\dimen1                        
      \rlap{\hbox to \dimen0{\hfil/\hfil}}      
      #1                                        
   \else                                        
      \rlap{\hbox to \dimen1{\hfil$#1$\hfil}}   
      /                                         
   \fi}                                         %

\begin{document}
\bibliographystyle{apsrev}

\preprint{BUHEP-01-25}

\title{Limits on the Mass of a Composite Higgs Boson: \\
an Update}


\author{R. Sekhar Chivukula}
\email{sekhar@bu.edu}
\affiliation{Physics Department, Boston University, 590 Commonwealth Ave\\
Boston, MA 02215 USA}

\author{Christian H\"olbling}
\email{christian.hoelbling@desy.de}
\affiliation{John von Neumann Institute of Computing (NIC), DESY Zeuthen \\
Platanenallee 6, D-15738 Zeuthen Germany}

\date{\today}

\begin{abstract}
  We discuss the bound on the mass of the Higgs boson arising from precision
  electroweak measurements in the context of the triviality of the scalar
  Higgs model. We show that, including possible effects from the underlying
  nontrivial dynamics, a Higgs boson mass of up to 500 GeV is consistent with
  current data.
\end{abstract}

\maketitle


\section{Introduction: Triviality
  of the Standard Higgs Model}

Current results from the LEP Electroweak Working Group \cite{Mele:2001} favor
a Higgs boson mass that is relatively light. The ``best-fit'' value
\footnote{This value for the Higgs mass arises from using the value
  $\Delta\alpha^{(5)}_{had}= 0.02738 \pm 0.00020$ for the contribution to the
  running of $\alpha_{em}$ from hadrons.}  for the Higgs mass is 106 GeV,
somewhat less than experimental lower bound \cite{Lhwg:2001} of 114.1 GeV (at
95\% confidence level).  The 95\% CL upper bound from precision measurements,
in the context of the standard model, is 222 GeV.  It is possible that, as
these data suggest, the Higgs boson lies around the corner and will be
discovered at relatively low masses. On the other hand, it is important to
consider alternatives and to understand what class of models can be
consistent with precision electroweak tests.  In this talk, we will show that
even minor modifications to the standard electroweak theory allow for a
substantially heavier Higgs boson\footnote{For a more complete discussion,
  see \protect\cite{Chivukula:2000px} and references therein.}.

This task is motivated by the fact that the standard one-doublet Higgs model
{\it does not strictly exist} as a continuum field theory
\cite{Wilson:1971bg,Wilson:1971dh,Wilson:1974jj}. This is because the
$\beta$-function for the Higgs-boson self-coupling is {\it positive}. For any
finite low-energy coupling, the running coupling-constant has a Landau pole:
it diverges at some finite energy.  Conversely, defining the model in terms
of a momentum-space cutoff $\Lambda$, the continuum limit is found by taking
$\Lambda \to \infty$ while holding all low-energy properties fixed. In this
limit, one finds that $\lambda \to 0$ --- {\it i.e.} the only continuum limit
is free or trivial.

The triviality of the scalar sector of the standard one-doublet Higgs model
implies that this theory is only an effective low-energy theory valid below
some finite cut-off scale $\Lambda$.  Given a value of {$m^2_H = 2
  \lambda(m_H) v^2$}, there is an {\it upper} bound on {$\Lambda$}.  An {\it
  estimate} of this bound \cite{Dashen:1983ts} can be obtained by integrating
the {one-loop} $\beta$-function, which yields
\beq
\Lambda \laem m_H \exp\left({4\pi^2v^2\over 3m^2_H}\right)~.
\label{landau}
\eeq
For a light Higgs, the bound above is at uninterestingly high scales
and the effects of the underlying dynamics can be too small to be
phenomenologically relevant. For a Higgs mass of order a few hundred
GeV, however, effects from the underlying physics can become
important. We will refer to these theories generically as ``composite
Higgs'' models.

\section{$T$, $S$, and $U$ in Composite Higgs Models}

\begin{figure}[htb]
\begin{center}
\begin{minipage}[t]{7.5cm}
\includegraphics[width=7.5cm]{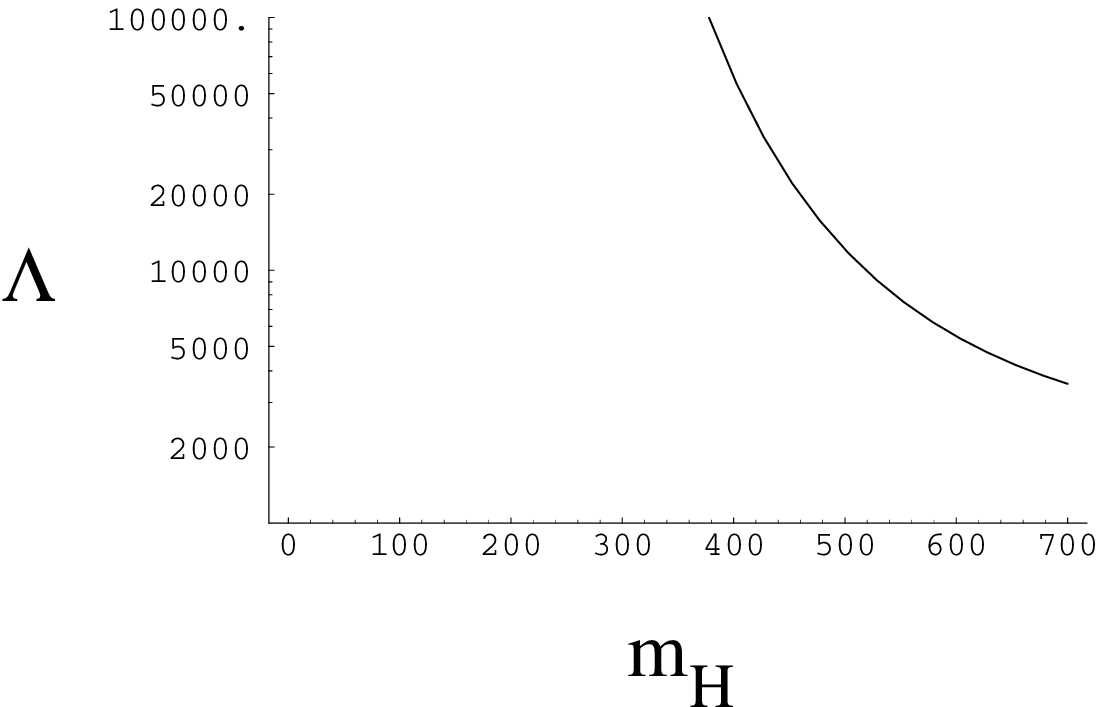}
\caption{Upper bound on scale $\Lambda$ as per eqn. (\protect\ref{landau}).}
\label{landaugraph}
\end{minipage}
\hspace{2mm}
\begin{minipage}[t]{7.5cm}
\includegraphics[width=7.5cm]{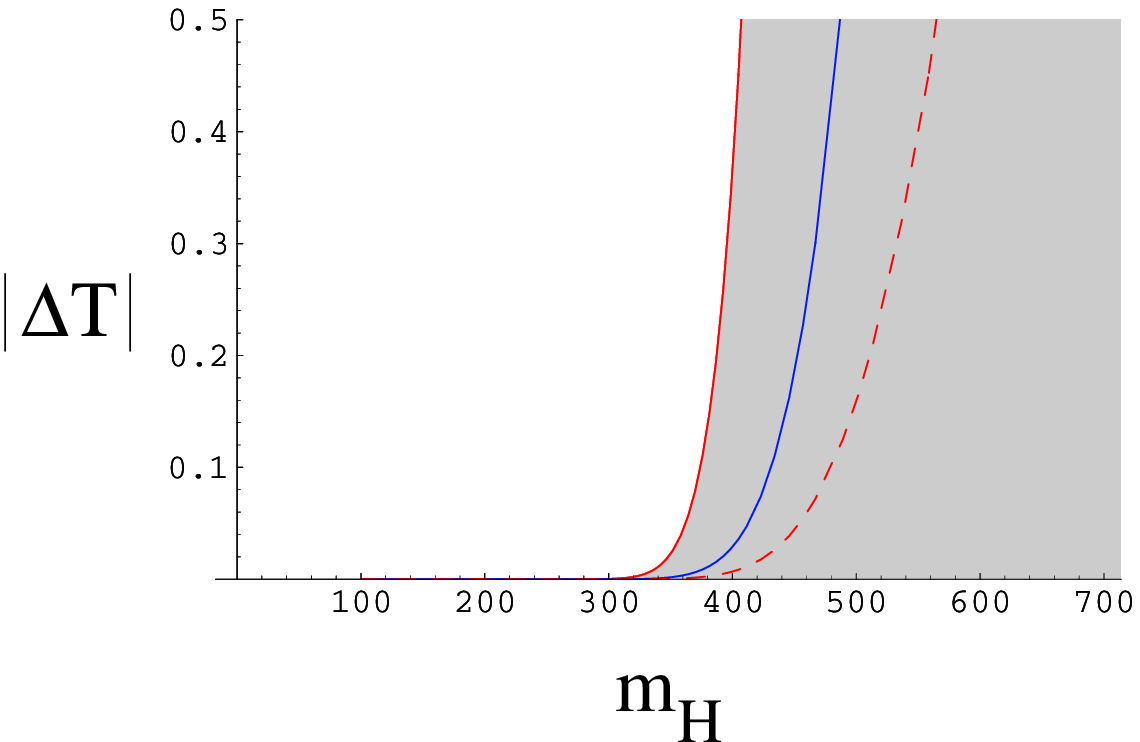}
\caption{Lower bound on expected size of $|\Delta T|$ as per eqn. (\protect\ref{tbound}),
for $|b|\kappa^2=16\pi^2$, $4\pi$, and 3.}
\label{tboundgraph}
\end{minipage}
\end{center}
\end{figure}

In an $SU(2)_W \times U(1)_Y$ invariant scalar theory of a single doublet,
all interactions of dimension less than or equal to four also respect a
larger ``custodial'' symmetry \cite{Weinstein:1973gj,Sikivie:1980hm} which
insures the tree-level relation $\rho=M^2_W / M^2_Z \cos^2\theta_W\equiv 1$.
The leading custodial-symmetry violating operator is of dimension
six \cite{Buchmuller:1986jz,Grinstein:1991cd} and involves four Higgs
doublet fields $\phi$. In general, the underlying theory does not respect the
larger custodial symmetry, and we expect the interaction
\beq
{\lower35pt\hbox{\epsfysize=1.00 truein \epsfbox{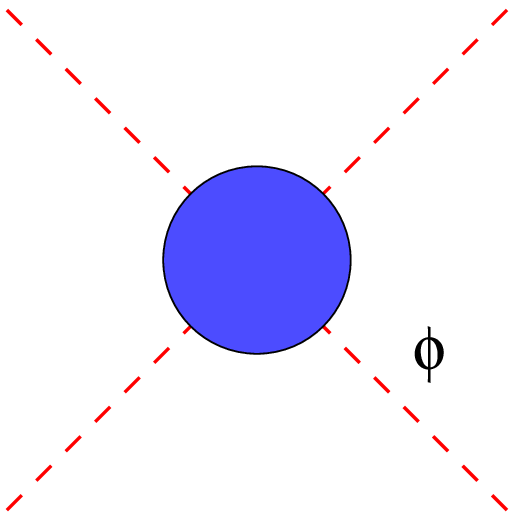}}}
\Rightarrow
{b { \kappa}^2 \over 2!\, { \Lambda}^2} 
(\phi^\dagger \stackrel{\leftrightarrow}{D^\mu} \phi)^2~,
\label{toperator}
\eeq
to appear in the low-energy effective theory.  Here $b$ is an unknown
coefficient of ${\cal O}(1)$, and ${\kappa}$ measures size of couplings of
the composite Higgs field. In a strongly-interacting theory, $\kappa$ is
expected \cite{Manohar:1984md,Georgi:1993dw} to be of ${\cal O}(4\pi)$.

Deviations in the low-energy theory from the standard model can be summarized
in terms of the ``oblique'' parameters
\cite{Peskin:1990zt,Peskin:1992sw,Golden:1991ig,Holdom:1990tc,Dobado:1991zh}
$S$, $T$, and $U$.  The operator in eqn. \ref{toperator} will give rise to a
deviation ($\Delta \rho= \varepsilon_1 = \alpha T$)
\beq { |\Delta T|} = { |b| {\kappa}^2 {v^2 \over \alpha(M_Z)
    { \Lambda}^2}} { \gaem} {|b|\kappa^2\, v^2 \over \alpha(M^2_Z)\, m^2_H}\,
\exp\left({-\,{8 \pi^2 v^2\over 3 m^2_H}}\right) ~, 
\label{tbound}
\eeq
where $v \approx 246$ GeV and we have used eqn. \ref{landau} to obtain
the final inequality.  The consequences of eqns. (\ref{landau}) and
(\ref{tbound}) are summarized in Figures \ref{landaugraph} and
\ref{tboundgraph}. The larger $m_H$, the lower $\Lambda$ and the
larger the expected value of $\Delta T$.  Current limits imply $|T|
\stackrel{<}{\sim} 0.5$, and hence ${ \Lambda \stackrel{>}{\sim} 4\,
  {\rm TeV} \cdot \kappa}$.  (For $\kappa \simeq 4\pi$, {$m_H \laem
  450$ GeV}.)

By contrast, the leading contribution to $S$ arises from
\beq
{\lower35pt\hbox{\epsfysize=1.00 truein \epsfbox{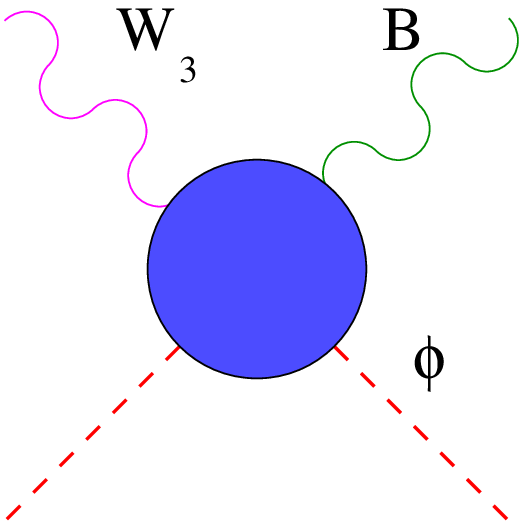}}}
\Rightarrow
-\,{a \over 2!\,{ \Lambda}^2} \left\{ [D_\mu,D_\nu] \phi \right\}^\dagger[D^\mu,D^\nu]\phi~.
\eeq
This gives rise to ($\varepsilon_3={\alpha S/4\sin^2\theta_W}$)
\beq \Delta S = {4 \pi a  v^2\over  { \Lambda}^2}~. \eeq
It is important to note that the size of contributions to $\Delta T$
and $\Delta S$ are very different
\beq 
{\Delta S \over \Delta T} = {a \over b} \left({4\pi \alpha \over { \kappa}^2}\right) ={\cal
  O}\left({10^{-1}\over { \kappa}^2}\right)~.
\eeq
Even for ${\kappa}\simeq 1$, $|\Delta S| \ll |\Delta T|$.

Finally, contributions to $U$ ($\varepsilon_2=-{\alpha U\over 4\sin^2\theta_W}$),
arise from
\beq
{c g^2 {\kappa^2}\over \Lambda^4} (\phi^\dagger W^{\mu\nu}\phi)^2
\eeq
and, being suppressed by $\Lambda^4$,  are typically much smaller than $\Delta T$.

\section{Limits on a Composite Higgs Boson}

\begin{figure}[htb]
\begin{center}
\includegraphics[width=12cm]{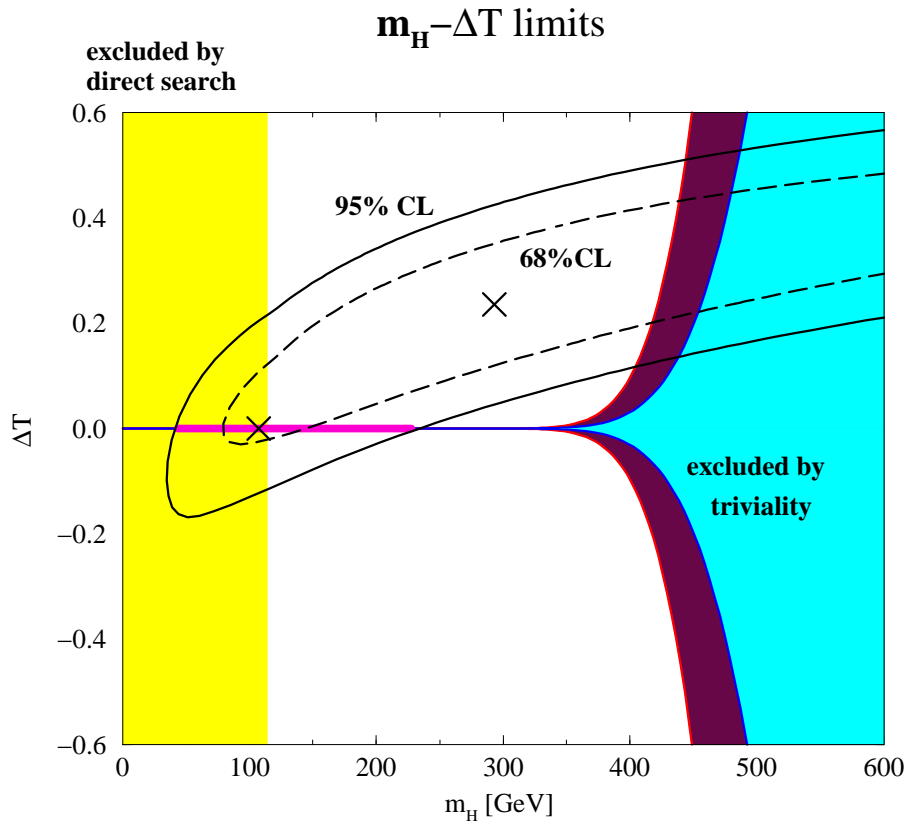}
\caption{68\% and 95\% CL regions allowed \protect\cite{Chivukula:2000px} in 
  $(m_H,\Delta T)$ plane by precision electroweak data
  \protect\cite{Mele:2001}. Fit allows for $m_t$, $\alpha_s$, and
  $\alpha_{em}$ to vary consistent with current limits
  \protect\cite{Chivukula:2000px}. Also shown by the the thick line on the
  $\Delta T=0$ axis is the usual one-dimensional 95\% CL limit quoted on the
  Higgs boson mass in the standard model, and the corresponding best fit.
  The triviality bound curves are for $|b|\kappa^2=4\pi$ and $4\pi^2$,
  corresponding to representative models \protect\cite{Chivukula:2000px}}.
\label{mht_zfitter}
\end{center}
\end{figure}

From triviality, we see that the Higgs model can only be an effective
theory valid below some high-energy scale $\Lambda$.  As the Higgs
becomes heavier, the scale $\Lambda$ {\it decreases}. Hence, the
expected size of contributions to $T$ {\it grow}, and are larger than
the expected contribution to $S$ or $U$. The limits from precision
electroweak data in $(m_H, \Delta T)$ plane shown in Figure
\ref{mht_zfitter}.  We see that, for positive $\Delta T$ at 95\% CL,
the allowed values of Higgs mass extend to well beyond 800 GeV. On the
other hand, not all values can be realized consistent with the bound
given in eqn.  (\ref{landau}). As shown in figure \ref{mht_zfitter},
values of Higgs mass beyond approximately 500 GeV would likely require
values of $\Delta T$ much larger than allowed by current measurements.

We should emphasize that these estimates are based on dimensional
arguments, and we are  not arguing that it is {\it impossible} to
construct a composite Higgs model consistent with precision
electroweak tests with $m_H$ greater than 500 GeV.  Rather, barring
accidental cancellations in a theory without a custodial symmetry,
contributions to $\Delta T$ consistent with eqn. \ref{landau} are
generally to be expected. 

\begin{figure}[htb]
\begin{center}
\includegraphics[width=12cm]{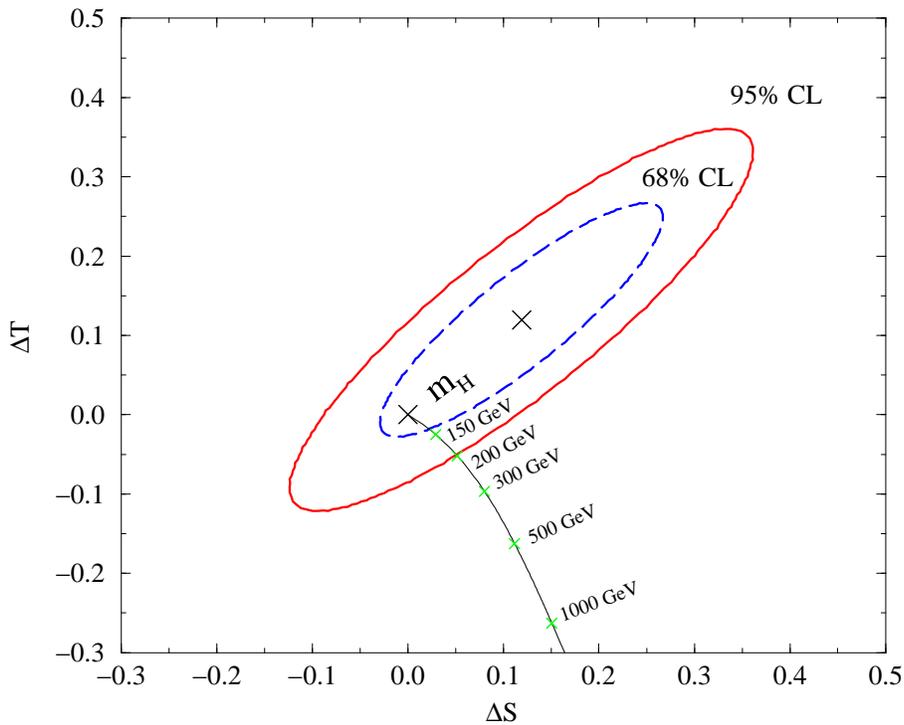}
\caption{68\% and 95\% CL regions allowed in $(\Delta S, \Delta T)$
plane by precision electroweak data \protect\cite{Mele:2001}.
Fit allows for $m_t$, $\alpha_s$, and $\alpha_{em}$ to vary consistent
with current limits \protect\cite{Chivukula:2000px}. Standard model prediction for
varying Higgs boson mass shown as parametric curve, with $m_H$ varying from
84 to 1000 GeV.}
\label{stplot_varymt}
\end{center}
\end{figure}

These results may also be understood by considering limits in the
$(S,T)$ plane for {\it fixed} $(m_H,m_t)$. In Figure
\ref{stplot_varymt}, changes from the nominal standard model best fit
($m_H=84$ GeV) value of the Higgs mass are displayed as contributions
to $\Delta S(m_H)$ and $\Delta T(m_H)$. Also shown are the 68\% and
95\% CL bounds on $\Delta S$ and $\Delta T$ consistent with current
data. We see that, for $m_H$ greater than ${\cal O}$(200 GeV), a
positive contribution to $T$ can bring the model within the allowed
region.

\section{The Top Quark Seesaw Model}

\begin{figure}[htb]
\begin{center}
\includegraphics*[bb=5 30 410 350,width=12cm]{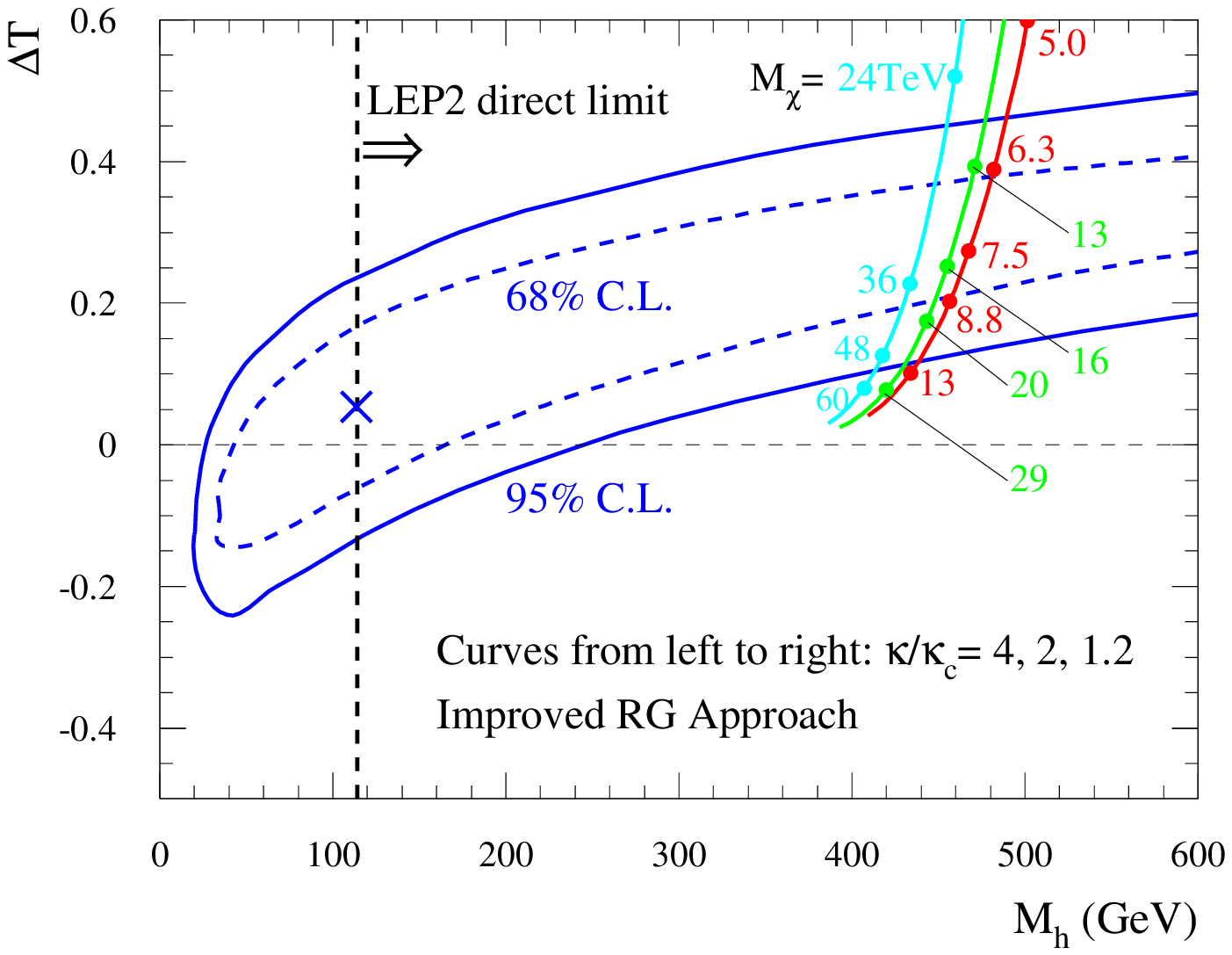}
\caption{$\Delta T$ vs. $m_H$ for the top-quark seesaw model plotted
  for various values of the mass of the heavy singlet quark, $m_\chi$, and
  various values of the (strong) topcolor-coupling, $\kappa \propto
  g^2_{tc}$, superimposed on fit to summer 2000 electroweak precision data.
  Courtesy of Hong-Jian He, \cite{He2001:aa}.}
\label{hjhe_fig}
\end{center}
\end{figure}

The top-quark seesaw theory of electroweak symmetry breaking
\cite{Dobrescu:1998nm,Chivukula:1998wd,Collins:1999rz,He:2001fz} provides a
simple example of a model with a potentially heavy composite Higgs boson
consistent with electroweak data. In this case, electroweak symmetry breaking
is due to the condensation, driven by a strong topcolor \cite{Hill:1991at}
gauge interaction, of the left-handed top-quark with a new right-handed
singlet fermion $\chi$. Such an interaction gives rise to a composite Higgs
field at low energies, and the mass of the top-color gauge boson sets the
scale of the Landau pole $\Lambda$ \cite{Bardeen:1990ds}.  The weak singlet
$\chi_L$ and $t_R$ fields are introduced so that the $2\times 2$ mass matrix,
\beq
\left(
\begin{array}{cc}
0 & m_{t\chi} \\
m_{\chi t} & m_{\chi \chi}
\end{array}
\right)
\eeq
is of seesaw form ($m_{\chi\chi} \gg m_{t\chi},\, m_{\chi t}$) and has a
light eigenvalue corresponding to the observed top quark. The value of $m_{t
  \chi}$ is related to the weak scale, and its value is estimated to be 600
GeV \cite{Dobrescu:1998nm}.

The coupling of the top-quark to $\chi$ violates custodial symmetry
in the same way that the top-quark mass does in the standard model.
The leading contribution to $T$ from the underlying top seesaw physics
arises from contributions to $W$ and $Z$ vacuum polarization diagrams
involving the $\chi$. This contribution is positive and is calculated to be
\cite{Dobrescu:1998nm,Collins:1999rz,He:2001fz}
\beq
\Delta T = {N_c \over 16 \pi^2 \alpha_{em}(M^2_Z) }\, {m^4_{t\chi}\over m^2_{\chi\chi} v^2}
\approx {0.7\over \alpha_{em}} \left({\Lambda^2 \over m^2_{\chi\chi}}\right)\,
\left({v^2\over \Lambda^2}\right)~,
\eeq
which is of the form of eqn. \ref{toperator} with $b\kappa^2 \propto
(\Lambda/m_{\chi\chi})^2$.  Note that $\Lambda/m_{\chi\chi}$ {\it
  cannot} be small since top-color gauge interactions must drive
$t\chi$ chiral symmetry breaking. 

A recent detailed analysis of precision electroweak constraints
\cite{He:2001fz,He2001:aa}, taking into account the running of the Higgs
self-coupling below the compositeness scale, yields the results shown in
Figure \ref{hjhe_fig}. The results show that the top quark seesaw model
essentially saturates the bounds implied by the triviality curves plotted in
Figure \ref{mht_zfitter}.

\section{Conclusions}

In conclusion, the triviality of the Standard Higgs model implies that
it is at best a {low-energy effective} theory valid below a scale
{$\Lambda$} characteristic of nontrivial underlying dynamics.  As the
Higgs mass {increases}, the upper bound on the scale $\Lambda$
{decreases}. If the underlying dynamics does not respect a custodial
symmetry, it will give rise to {corrections to $T$ of order $\kappa^2
  v^2/\alpha \Lambda^2$}, while the contributions to $S$ and $U$ are
likely to be much smaller. For this reason, it is necessary to
consider limits on a Higgs boson in the {$(m_H,\Delta T)$} plane. In
doing so, we see that a Higgs mass {larger than 200 GeV is
  consistent} with precision electroweak tests {if} there is a
positive $\Delta T$. Absent a custodial symmetry, however, Higgs
masses {larger than $\simeq 500$ GeV are unlikely}: the scale of
underlying physics is so low that $\Delta T$ is likely to be too
large.

\section{Acknowledgments}

We thank Bogdan Dobrescu, Nick Evans, and E.~H.~Simmons for fruitful
collaborations, and Hong-Jian He for extensive discussions on electroweak
fits in composite Higgs models and for allowing us to use Figure
\ref{hjhe_fig}. This work was supported in part by the U.S.  Department of
Energy under grant DE-FG02-91ER40676.

\medskip

\bigskip
\bibliography{higgs_snow}

\begin{thebibliography}{25}
\expandafter\ifx\csname natexlab\endcsname\relax\def\natexlab#1{#1}\fi
\expandafter\ifx\csname bibnamefont\endcsname\relax
  \def\bibnamefont#1{#1}\fi
\expandafter\ifx\csname bibfnamefont\endcsname\relax
  \def\bibfnamefont#1{#1}\fi
\expandafter\ifx\csname citenamefont\endcsname\relax
  \def\citenamefont#1{#1}\fi
\expandafter\ifx\csname url\endcsname\relax
  \def\url#1{\texttt{#1}}\fi
\expandafter\ifx\csname urlprefix\endcsname\relax\def\urlprefix{URL }\fi
\providecommand{\bibinfo}[2]{#2}
\providecommand{\eprint}[2][]{\url{#2}}

\bibitem[{\citenamefont{Mele}(2001)}]{Mele:2001}
\bibinfo{author}{\bibfnamefont{S.}~\bibnamefont{Mele}}
  (\bibinfo{collaboration}{LEP Electroweak Working Group})
  (\bibinfo{year}{2001}), \bibinfo{note}{contributed to International
  Europhysics Conference on High- Energy Physics (HEP 2001), Budapest, Hungary,
  12-18 Jul 2001}.

\bibitem[{Lhw(2001)}]{Lhwg:2001}
  (\bibinfo{year}{2001}), \bibinfo{note}{cERN-EP/2001-055}.

\bibitem[{\citenamefont{Wilson}(1971{\natexlab{a}})}]{Wilson:1971bg}
\bibinfo{author}{\bibfnamefont{K.~G.} \bibnamefont{Wilson}},
  \bibinfo{journal}{Phys. Rev.} \textbf{\bibinfo{volume}{B4}},
  \bibinfo{pages}{3174} (\bibinfo{year}{1971}{\natexlab{a}}).

\bibitem[{\citenamefont{Wilson}(1971{\natexlab{b}})}]{Wilson:1971dh}
\bibinfo{author}{\bibfnamefont{K.~G.} \bibnamefont{Wilson}},
  \bibinfo{journal}{Phys. Rev.} \textbf{\bibinfo{volume}{B4}},
  \bibinfo{pages}{3184} (\bibinfo{year}{1971}{\natexlab{b}}).

\bibitem[{\citenamefont{Wilson and Kogut}(1974)}]{Wilson:1974jj}
\bibinfo{author}{\bibfnamefont{K.~G.} \bibnamefont{Wilson}} \bibnamefont{and}
  \bibinfo{author}{\bibfnamefont{J.~B.} \bibnamefont{Kogut}},
  \bibinfo{journal}{Phys. Rept.} \textbf{\bibinfo{volume}{12}},
  \bibinfo{pages}{75} (\bibinfo{year}{1974}).

\bibitem[{\citenamefont{Dashen and Neuberger}(1983)}]{Dashen:1983ts}
\bibinfo{author}{\bibfnamefont{R.}~\bibnamefont{Dashen}} \bibnamefont{and}
  \bibinfo{author}{\bibfnamefont{H.}~\bibnamefont{Neuberger}},
  \bibinfo{journal}{Phys. Rev. Lett.} \textbf{\bibinfo{volume}{50}},
  \bibinfo{pages}{1897} (\bibinfo{year}{1983}).

\bibitem[{\citenamefont{Weinstein}(1973)}]{Weinstein:1973gj}
\bibinfo{author}{\bibfnamefont{M.}~\bibnamefont{Weinstein}},
  \bibinfo{journal}{Phys. Rev.} \textbf{\bibinfo{volume}{D8}},
  \bibinfo{pages}{2511} (\bibinfo{year}{1973}).

\bibitem[{\citenamefont{Sikivie et~al.}(1980)\citenamefont{Sikivie, Susskind,
  Voloshin, and Zakharov}}]{Sikivie:1980hm}
\bibinfo{author}{\bibfnamefont{P.}~\bibnamefont{Sikivie}},
  \bibinfo{author}{\bibfnamefont{L.}~\bibnamefont{Susskind}},
  \bibinfo{author}{\bibfnamefont{M.~B.} \bibnamefont{Voloshin}},
  \bibnamefont{and} \bibinfo{author}{\bibfnamefont{V.}~\bibnamefont{Zakharov}},
  \bibinfo{journal}{Nucl. Phys.} \textbf{\bibinfo{volume}{B173}},
  \bibinfo{pages}{189} (\bibinfo{year}{1980}).

\bibitem[{\citenamefont{Buchmuller and Wyler}(1986)}]{Buchmuller:1986jz}
\bibinfo{author}{\bibfnamefont{W.}~\bibnamefont{Buchmuller}} \bibnamefont{and}
  \bibinfo{author}{\bibfnamefont{D.}~\bibnamefont{Wyler}},
  \bibinfo{journal}{Nucl. Phys.} \textbf{\bibinfo{volume}{B268}},
  \bibinfo{pages}{621} (\bibinfo{year}{1986}).

\bibitem[{\citenamefont{Grinstein and Wise}(1991)}]{Grinstein:1991cd}
\bibinfo{author}{\bibfnamefont{B.}~\bibnamefont{Grinstein}} \bibnamefont{and}
  \bibinfo{author}{\bibfnamefont{M.~B.} \bibnamefont{Wise}},
  \bibinfo{journal}{Phys. Lett.} \textbf{\bibinfo{volume}{B265}},
  \bibinfo{pages}{326} (\bibinfo{year}{1991}).

\bibitem[{\citenamefont{Manohar and Georgi}(1984)}]{Manohar:1984md}
\bibinfo{author}{\bibfnamefont{A.}~\bibnamefont{Manohar}} \bibnamefont{and}
  \bibinfo{author}{\bibfnamefont{H.}~\bibnamefont{Georgi}},
  \bibinfo{journal}{Nucl. Phys.} \textbf{\bibinfo{volume}{B234}},
  \bibinfo{pages}{189} (\bibinfo{year}{1984}).

\bibitem[{\citenamefont{Georgi}(1993)}]{Georgi:1993dw}
\bibinfo{author}{\bibfnamefont{H.}~\bibnamefont{Georgi}},
  \bibinfo{journal}{Phys. Lett.} \textbf{\bibinfo{volume}{B298}},
  \bibinfo{pages}{187} (\bibinfo{year}{1993}), \eprint{arXiv:hep-ph/9207278}.

\bibitem[{\citenamefont{Peskin and Takeuchi}(1990)}]{Peskin:1990zt}
\bibinfo{author}{\bibfnamefont{M.~E.} \bibnamefont{Peskin}} \bibnamefont{and}
  \bibinfo{author}{\bibfnamefont{T.}~\bibnamefont{Takeuchi}},
  \bibinfo{journal}{Phys. Rev. Lett.} \textbf{\bibinfo{volume}{65}},
  \bibinfo{pages}{964} (\bibinfo{year}{1990}).

\bibitem[{\citenamefont{Peskin and Takeuchi}(1992)}]{Peskin:1992sw}
\bibinfo{author}{\bibfnamefont{M.~E.} \bibnamefont{Peskin}} \bibnamefont{and}
  \bibinfo{author}{\bibfnamefont{T.}~\bibnamefont{Takeuchi}},
  \bibinfo{journal}{Phys. Rev.} \textbf{\bibinfo{volume}{D46}},
  \bibinfo{pages}{381} (\bibinfo{year}{1992}).

\bibitem[{\citenamefont{Golden and Randall}(1991)}]{Golden:1991ig}
\bibinfo{author}{\bibfnamefont{M.}~\bibnamefont{Golden}} \bibnamefont{and}
  \bibinfo{author}{\bibfnamefont{L.~J.} \bibnamefont{Randall}},
  \bibinfo{journal}{Nucl. Phys.} \textbf{\bibinfo{volume}{B361}},
  \bibinfo{pages}{3} (\bibinfo{year}{1991}).

\bibitem[{\citenamefont{Holdom and Terning}(1990)}]{Holdom:1990tc}
\bibinfo{author}{\bibfnamefont{B.}~\bibnamefont{Holdom}} \bibnamefont{and}
  \bibinfo{author}{\bibfnamefont{J.}~\bibnamefont{Terning}},
  \bibinfo{journal}{Phys. Lett.} \textbf{\bibinfo{volume}{B247}},
  \bibinfo{pages}{88} (\bibinfo{year}{1990}).

\bibitem[{\citenamefont{Dobado et~al.}(1991)\citenamefont{Dobado, Espriu, and
  Herrero}}]{Dobado:1991zh}
\bibinfo{author}{\bibfnamefont{A.}~\bibnamefont{Dobado}},
  \bibinfo{author}{\bibfnamefont{D.}~\bibnamefont{Espriu}}, \bibnamefont{and}
  \bibinfo{author}{\bibfnamefont{M.~J.} \bibnamefont{Herrero}},
  \bibinfo{journal}{Phys. Lett.} \textbf{\bibinfo{volume}{B255}},
  \bibinfo{pages}{405} (\bibinfo{year}{1991}).

\bibitem[{\citenamefont{Chivukula et~al.}(2000)\citenamefont{Chivukula,
  Holbling, and Evans}}]{Chivukula:2000px}
\bibinfo{author}{\bibfnamefont{R.~S.} \bibnamefont{Chivukula}},
  \bibinfo{author}{\bibfnamefont{C.}~\bibnamefont{Holbling}}, \bibnamefont{and}
  \bibinfo{author}{\bibfnamefont{N.}~\bibnamefont{Evans}},
  \bibinfo{journal}{Phys. Rev. Lett.} \textbf{\bibinfo{volume}{85}},
  \bibinfo{pages}{511} (\bibinfo{year}{2000}), \eprint{arXiv:hep-ph/0002022}.

\bibitem[{\citenamefont{He}(2001)}]{He2001:aa}
\bibinfo{author}{\bibfnamefont{H.-J.} \bibnamefont{He}}
  (\bibinfo{year}{2001}), \bibinfo{note}{private communication}.

\bibitem[{\citenamefont{Dobrescu and Hill}(1998)}]{Dobrescu:1998nm}
\bibinfo{author}{\bibfnamefont{B.~A.} \bibnamefont{Dobrescu}} \bibnamefont{and}
  \bibinfo{author}{\bibfnamefont{C.~T.} \bibnamefont{Hill}},
  \bibinfo{journal}{Phys. Rev. Lett.} \textbf{\bibinfo{volume}{81}},
  \bibinfo{pages}{2634} (\bibinfo{year}{1998}), \eprint{hep-ph/9712319}.

\bibitem[{\citenamefont{Chivukula et~al.}(1999)\citenamefont{Chivukula,
  Dobrescu, Georgi, and Hill}}]{Chivukula:1998wd}
\bibinfo{author}{\bibfnamefont{R.~S.} \bibnamefont{Chivukula}},
  \bibinfo{author}{\bibfnamefont{B.~A.} \bibnamefont{Dobrescu}},
  \bibinfo{author}{\bibfnamefont{H.}~\bibnamefont{Georgi}}, \bibnamefont{and}
  \bibinfo{author}{\bibfnamefont{C.~T.} \bibnamefont{Hill}},
  \bibinfo{journal}{Phys. Rev.} \textbf{\bibinfo{volume}{D59}},
  \bibinfo{pages}{075003} (\bibinfo{year}{1999}), \eprint{hep-ph/9809470}.

\bibitem[{\citenamefont{Collins et~al.}(2000)\citenamefont{Collins, Grant, and
  Georgi}}]{Collins:1999rz}
\bibinfo{author}{\bibfnamefont{H.}~\bibnamefont{Collins}},
  \bibinfo{author}{\bibfnamefont{A.}~\bibnamefont{Grant}}, \bibnamefont{and}
  \bibinfo{author}{\bibfnamefont{H.}~\bibnamefont{Georgi}},
  \bibinfo{journal}{Phys. Rev.} \textbf{\bibinfo{volume}{D61}},
  \bibinfo{pages}{055002} (\bibinfo{year}{2000}), \eprint{hep-ph/9908330}.

\bibitem[{\citenamefont{He et~al.}(2001)\citenamefont{He, Hill, and
  Tait}}]{He:2001fz}
\bibinfo{author}{\bibfnamefont{H.-J.} \bibnamefont{He}},
  \bibinfo{author}{\bibfnamefont{C.~T.} \bibnamefont{Hill}}, \bibnamefont{and}
  \bibinfo{author}{\bibfnamefont{T.~M.~P.} \bibnamefont{Tait}}
  (\bibinfo{year}{2001}), \eprint{arXiv:hep-ph/0108041}.

\bibitem[{\citenamefont{Hill}(1991)}]{Hill:1991at}
\bibinfo{author}{\bibfnamefont{C.~T.} \bibnamefont{Hill}},
  \bibinfo{journal}{Phys. Lett.} \textbf{\bibinfo{volume}{B266}},
  \bibinfo{pages}{419} (\bibinfo{year}{1991}).

\bibitem[{\citenamefont{Bardeen et~al.}(1990)\citenamefont{Bardeen, Hill, and
  Lindner}}]{Bardeen:1990ds}
\bibinfo{author}{\bibfnamefont{W.~A.} \bibnamefont{Bardeen}},
  \bibinfo{author}{\bibfnamefont{C.~T.} \bibnamefont{Hill}}, \bibnamefont{and}
  \bibinfo{author}{\bibfnamefont{M.}~\bibnamefont{Lindner}},
  \bibinfo{journal}{Phys. Rev.} \textbf{\bibinfo{volume}{D41}},
  \bibinfo{pages}{1647} (\bibinfo{year}{1990}).

\end{thebibliography}

\end{document}